\documentclass[nofootinbib,preprint,preprintnumbers,a4paper,11pt]{revtex4}
\usepackage{epsfig}
\usepackage{footnote}
\usepackage{ulem}
\usepackage{color}
\usepackage{array}
\usepackage{amssymb}
\usepackage{amsmath}
\usepackage{graphicx}
\usepackage{subfigure}
\usepackage{longtable}
\usepackage{verbatim}
\usepackage{amsfonts}
\usepackage{hyperref}
\usepackage{multirow}
\usepackage{cancel}

\begin{document}
\title{Pursuit of new physics within the $\Lambda_c\to \Delta K$ decays}
\author{Di Wang}
\email{wangdi@hunnu.edu.cn}
\affiliation{Department of Physics, Hunan Normal University, and Key Laboratory of Low-Dimensional Quantum Structures and Quantum Control of Ministry of Education, Changsha 410081, China}

\begin{abstract}

We point out the $CP$ asymmetry in the chain decay of $\Lambda_c^+\to \Delta^+K(t)(\to \pi^+\pi^-)$ is determined by the branching fractions of $\Lambda^+_c\to \Delta^{++} K^-$, $\Lambda^+_c\to \Delta^{+} K^0_{S,L}$ and $\Lambda^+_c\to \Delta^{0} K^+$ modes under the isospin symmetry.
The ambiguities from loop-induced quantities and $SU(3)_F$ breaking hadronic effects are avoided.
Once the $CP$ asymmetry in the $\Lambda_c^+$ decays into $\Delta^+$ and neutral kaons is confirmed by experiments, we can check if it is beyond the Standard Model, or verify the $CP$-violating effect resulted from the interference between the Cabibbo-favored and the doubly Cabibbo-suppressed amplitudes with the neutral kaon mixing.
Future measurements of branching fractions play a critical role in reducing the uncertainties.

\end{abstract}

\maketitle

\section{Introduction}
$CP$ asymmetry in heavy quark weak decay provides a window to test the Standard Model (SM) and search for new physics (NP).
Charmed hadron system is the only platform to probe $CP$ asymmetry in the up-type quark decay in hadron.
The LHCb collaboration reported the discovery of $CP$ asymmetry in the charm meson decays in 2019 \cite{Aaij:2019kcg},
\begin{align}\label{z1}
  \Delta A_{CP} &\equiv A_{CP}(D^0\to K^+K^-)-A_{CP}(D^0\to \pi^+\pi^-)\nonumber\\&\quad=(-1.54\pm0.29)\times 10^{-3}.
\end{align}
And the $CP$ asymmetries of single decay modes were measured recently \cite{LHCb:2022vcc}.
In theoretical aspect, the ambiguity of penguin topology results in great difficulty in evaluating $CP$ asymmetry of the singly Cabibbo-suppressed (SCS) decay.
The Quantum Chromodynamics (QCD) inspired approaches do not work well in charm scale.
Damaged by the almost exact cancellation between the Cabibbo-Kobayashi-Maskawa (CKM) matrix elements $V_{cd}^*V_{ud}$ and $V_{cs}^*V_{us}$, the penguin topologies fail to extract from branching fractions.
There are two controversial viewpoints for the observed $CP$ asymmetry in charm in literature, regarding it as a signal of new physics \cite{Buras:2021rdg,Calibbi:2019bay,Chala:2019fdb,Dery:2019ysp} or the non-perturbative QCD enhancements to penguin \cite{Wang:2022nbm,Bediaga:2022sxw,Wang:2021rhd,Schacht:2021jaz,Li:2019hho,Cheng:2019ggx,Grossman:2019xcj}.

Two sum rules of direct $CP$ asymmetry in the linear $SU(3)_F$ breaking were proposed in Ref.~\cite{Muller:2015rna} for avoiding predicting individual $CP$ asymmetries.
In these sum rules, the primary uncertainties induced by penguin topologies are eliminated.
To uncover new physics, some coefficients of the sum rules, including the strong phases of tree topologies, need to be determined by the global fit of branching fractions.
However, there are too many parameters in the global fit \cite{Muller:2015lua}.  The uncertainties are hard to reduce.
Moreover, the $CP$ asymmetry sum rules are broken by the second-order $SU(3)_F$ breaking effects which turn out to be non-negligible in charm decay \cite{HFLAV:2019otj,Gronau:2015rda,Gronau:2013xba,Gronau:2012kq}.

For the purpose of circumventing penguin topology, the $CP$ asymmetry of $D^0\to K^0_SK^0_S$ was analyzed in Ref.~\cite{Nierste:2015zra}.  The upper bound is given by $|A_{CP}^{\rm dir}|\leq 1.1\%\,\,(95\%\,\,\rm C.L.)$.
If the future data exceed the upper bound, it might be a signal of new physics.
But a QCD enhancement of the penguin annihilation $PA$ cannot be excluded \cite{Nierste:2015zra}.
Besides, the theoretical uncertainties cannot be well controlled since the tree-level decay amplitudes are arisen from the $SU(3)_F$ breaking.
The similar problem also appears in the $D^0\to K^0_SK^{*0}$ and $D^0\to K^0_S\overline K^{*0}$ modes.
The reliability of extracted $E_P$ and $E_V$ is suffered from the uncontrollable $SU(3)_F$ breaking effects \cite{Nierste:2017cua,Cheng:2021yrn}.
And a large penguin annihilation cannot be excluded either.

$CP$ asymmetry also appears in the Cabibbo-favored (CF) and doubly Cabibbo-suppressed (DCS) charmed hadron decays into neutral kaons \cite{Lipkin:1999qz,DAmbrosio:2001mpr,Bigi:1994aw,Xing:1995jg,Bianco:2003vb,Grossman:2011zk,Belle:2012ygt}.
The time-dependent and time-integrated $CP$ asymmetries in the chain decays of $D^{+}\to \pi^{+}K(t)(\to \pi^+\pi^-)$ and $D_{s}^{+}\to K^{+}K(t)(\to \pi^+\pi^-)$, where $K(t)$ represents a time-evolved neutral kaon $K^0(t)$ or $\overline K^0(t)$ with $t$ being the time difference between the charm decays and the neutral kaon decays in the kaon rest frame, were studied in Ref.~\cite{Yu:2017oky}.
A new $CP$-violating effect resulted from the interference between the DCS and CF amplitudes with the mixing of final-state neutral kaons was reported. The total $CP$ asymmetries were estimated by the so-called factorization-assisted topological-amplitude (FAT) approach \cite{FAT}.
Unfortunately, the FAT approach cannot precisely describe the $SU(3)_F$ breaking effects of $D$ meson decays \cite{BESIII:2022xhe,Wang:2017ksn}.
The estimated strong phases and $CP$ asymmetries are not convincing.

Another window to search for new physics in charm is the neutral $D$ meson mixing.
Attributed to the charm scale being too heavy to apply the chiral perturbation theory and too light to apply the heavy quark expansion, and the deep suppression from the GIM mechanism, the theoretical evaluation of $D$ mixing system is a challenging subject \cite{Li:2022jxc,Jubb:2016mvq,Falk:2001hx,Bigi:2000wn,Lenz:2020efu,Li:2020xrz,Jiang:2017zwr,Cheng:2010rv,Datta:1984jx,Georgi:1992as,Ohl:1992sr,Golowich:2005pt}.

To establish a "smoking gun" signal of new physics in charm, one need reliable SM predictions.
With the helplessness of controlling the theoretical uncertainties of charmed meson system, we turn to study the charmed baryon decay. The $CP$ asymmetries of charmed baryon decays into neutral kaons such as $\Lambda^+_c\to pK(t)(\to \pi^+\pi^-)$ were analyzed in our previous work \cite{Wang:2017gxe}. Just like the case of $D$ decay, the large uncertainties induced by the $SU(3)_F$ breaking effects cannot be removed.
Contrary to the pseudoscalar/vector meson and octet baryon, the flavor symmetry of decuplet baryon would help us to eliminate some theoretical uncertainties.
In this work, we investigate $CP$ asymmetry in the $\Lambda_c^+$ decaying into $\Delta^+$ and neutral kaons.
It is found the hadronic parameters can be quantified by the branching fractions of several $\Lambda_c\to \Delta K$ modes in the isospin symmetry without \textit{ad hoc} assumptions.
Once the $CP$ asymmetry of $\Lambda^+_c\to \Delta^{+}K(t)(\to \pi^+\pi^-)$ is confirmed by experiments, we can check whether it is beyond the Standard Model or not.

The rest of this paper is structured as follows.
In Sec. \ref{cpv}, we discuss the $CP$ asymmetry of $\Lambda^+_c\to \Delta^{+}K(t)(\to \pi^+\pi^-)$.
In Sec. \ref{ext}, we show how to extract the theoretical parameters that determine $CP$ asymmetry.
And Sec. \ref{sum} is a brief summary. The isospin analysis of the $\Lambda_c\to \Delta K$ modes is listed in Appendix. \ref{iso}.

\section{$CP$ asymmetry}\label{cpv}
In this section, we discuss $CP$ asymmetry in the $\Lambda^+_c\to \Delta^{+}K(t)(\to \pi^+\pi^-)$ mode.
The mass eigenstates of neutral kaons $K_S^0$ and $K_L^0$ are linear combinations of the flavor
eigenstates $K^0$ and $\overline K^0$,
 \begin{equation}\label{eq:KSKL}
|K_{S,L}^0\rangle  =   \frac{1+\epsilon}{\sqrt{2(1+|\epsilon|^2)}}|K^0\rangle\mp
\frac{1-\epsilon}{\sqrt{2(1+|\epsilon|^2)}}|\overline{K}^0\rangle,
 \end{equation}
where $\epsilon$ is a complex parameter characterizing the $CP$ asymmrtry in the kaon mixing with
$|\epsilon|=(2.228\pm0.011)\times10^{-3}$ and $\phi_{\epsilon}=43.52\pm0.05^{\circ}$ \cite{PDG}.
In experiments, a $K^0_S$ candidate is reconstructed by its decay into two charged pions at a time close to its lifetime. Not only $K^0_S$, but also $K^0_L$ serve as the intermediate states in the $\Lambda^+_c\to \Delta^{+}K(t)(\to \pi^+\pi^-)$ mode through the $K^0_S-K^0_L$ oscillation \cite{Grossman:2011zk}.
The time-dependent $CP$ asymmetry in the $\Lambda^+_c\to \Delta^{+}K(t)(\to \pi^+\pi^-)$ mode is defined by
\begin{equation}\label{m1}
A_{CP}(t) \equiv\frac{\Gamma_{\pi\pi}(t)-\overline
\Gamma_{\pi\pi}(t)}{\Gamma_{\pi\pi}(t)+\overline\Gamma_{\pi\pi}(t)},
\end{equation}
where
\begin{align}
  \Gamma_{\pi\pi}(t)&\equiv\Gamma(\Lambda^+_c\to \Delta^+K(t)(\to \pi^{+}\pi^{-})), \nonumber\\
\overline\Gamma_{\pi\pi}(t)&\equiv\Gamma(\Lambda^-_c\to \Delta^-K(t)
(\to \pi^{+}\pi^{-})).
\end{align}
We write the ratio between $\mathcal{A}(\Lambda^+_c\to \Delta^{+} K^0)$ and $\mathcal{A}(\Lambda^+_c\to \Delta^{+}\overline K^0)$ as
\begin{equation}\label{r}
\mathcal{A}(\Lambda^+_c\to \Delta^{+} K^0)/\mathcal{A}(\Lambda^+_c\to \Delta^{+}\overline K^0) = r\,e^{i(\phi+\delta)},
\end{equation}
with the magnitude $r$, the relative strong phase $\delta$, and the weak phase $\phi=(-6.2\pm 0.4)\times 10^{-4}$ in the SM \cite{PDG}.

Similarly to the $D$ meson decay, the time-dependent $CP$ asymmetry of $\Lambda^+_c\to \Delta^{+}K(t)(\to \pi^+\pi^-)$ decay is derived to be
\begin{align}\label{eq:ACPt}
 A_{CP}(t)\simeq
\left(A_{CP}^{\overline K^0}(t)+A_{CP}^{\rm dir}(t)+A_{CP}^{\rm int}(t)\right)/{D(t)},
\end{align}
with
\begin{align}\label{eq:AcpK0}
A_{CP}^{\overline K^0}(t)
&=2e^{-\Gamma_{K^0_S}t}  \mathcal{R}e(\epsilon)-2e^{-\Gamma_K t}
\Big(\mathcal{R}e(\epsilon)\cos(\Delta m_Kt)
\nonumber\\&\qquad\qquad+\mathcal{I}m(\epsilon)\sin(\Delta m_Kt)\Big),\\
\label{eq:Acpint}
A_{CP}^{\rm int}(t)
&=-4\,r\cos\phi\sin\delta\Big[e^{-\Gamma_{K^0_S}t}\mathcal{I}m(\epsilon)-e^{-\Gamma_K t}
\nonumber\\&\qquad\qquad\times\Big(\mathcal{I}m(\epsilon)\cos(\Delta m_Kt)-\mathcal{R}e(\epsilon)\sin(\Delta m_Kt)\Big)\Big],\\
A_{CP}^{\rm dir}(t)&=e^{-\Gamma_{K^0_S}t}\,2\,r\sin\delta\sin\phi,\\
D(t)&= e^{-\Gamma_{K^0_S}t}(1-2\,r\cos\delta\cos\phi),
\end{align}
where the average of widths is $\Gamma_K\equiv(\Gamma_{K^0_S}+\Gamma_{K^0_L})/2$, and the differences of widths and masses are $\Delta\Gamma_K\equiv\Gamma_{K^0_S}-\Gamma_{K^0_L}$ and $\Delta m_K\equiv m_{K^0_L}-m_{K^0_S}$, respectively.
The first term in Eq.~\eqref{eq:ACPt}, which is independent of the hadronic parameters $r$ and $\delta$, is $CP$ asymmetry in the neutral kaon mixing \cite{Grossman:2011zk}.
The second term is direct $CP$ asymmetry induced by the interference between the tree level CF and DCS amplitudes.
The third term is the interference between the CF and DCS amplitudes with the neutral kaon mixing, a new $CP$-violating effect pointed out in Ref.~\cite{Yu:2017oky}.
Measurements of $CP$ asymmetries depend on time intervals selected in experiments.
The time-integrated $CP$ asymmetry in the limit of $t_1\ll \tau_S \ll t_2 \ll \tau_L$ is
\begin{align}\label{eq:t1t2limit}
 &A_{CP}(t_1\ll \tau_S\ll t_2 \ll \tau_L)
\simeq\left(A_{CP}^{\overline K^0}+A_{CP}^{\rm dir}+A_{CP}^{\rm int}\right)/{D}\nonumber\\~~~~
& ~~~~~=\frac{-2\mathcal{R}e(\epsilon)
+2\,r\sin\phi\sin\delta-4\,\mathcal{I}m(\epsilon)\,r\cos\phi\sin\delta}
 {1-2\,r\cos\phi\cos\delta}.
\end{align}
In the SM, $A_{CP}^{\overline K^0}$ is well determined by parameter $\epsilon$.
$A_{CP}^{\rm int}$ and $A_{CP}^{\rm dir}$ are estimated to be $\mathcal{O}(10^{-4})$ and $\mathcal{O}(10^{-5})$, respectively.

According to Eqs.~\eqref{eq:ACPt}$\sim$\eqref{eq:t1t2limit}, there are two hadronic inputs in evaluating the time-dependent and time-integrated $CP$ asymmetries, $r$ and $\delta$.
If $r$ and $\delta$ are well determined from the branching fractions, the theoretical prediction for the $CP$ asymmetry in the $\Lambda^+_c\to \Delta^{+}K(t)(\to \pi^+\pi^-)$ mode will be very precise.
On the other hand, if there exists a beyond-SM weak phase in the DCS or CF transitions, the $CP$ asymmetry will differ from the SM prediction.
Unlike the pure CF or DCS modes or the semileptonic modes, the relative strong phase between $\Lambda^+_c\to \Delta^+K^0 $ and $\Lambda^+_c\to \Delta^+\overline K^0 $ possibly serves as the strong phase between the SM amplitude and the NP amplitude.
It gets rid of the potential scenario that ultraviolet new physics is hidden due to absence of the relative strong phase.

\section{Extraction of hadronic parameters}\label{ext}
\begin{figure}[!tph]
\centering
\includegraphics[width=0.48\textwidth]{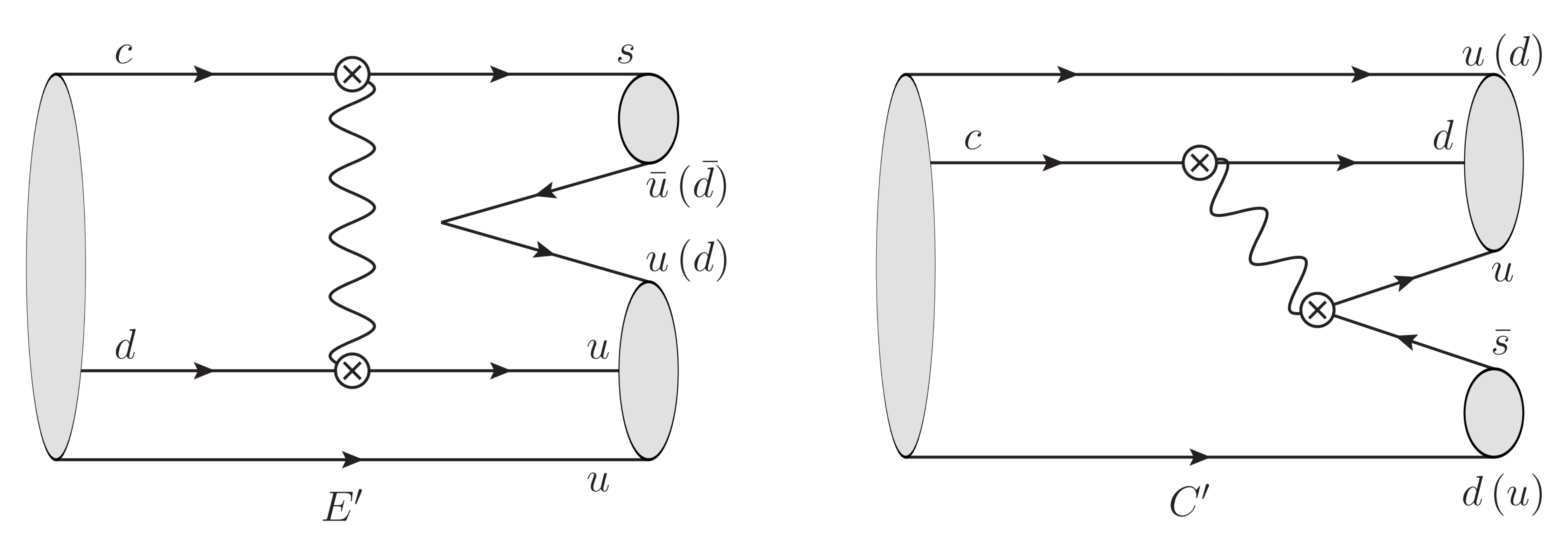}
\caption{Topological diagrams contributing to the $\Lambda^+_c\to \Delta^{++}K^-$ ($\Lambda^+_c\to \Delta^+\overline K^0$) and $\Lambda^+_c\to \Delta^+K^0$ ($\Lambda^+_c\to \Delta^0K^+$) modes respectively.}
\label{topo}
\end{figure}

In the $\Lambda_c\to \Delta K$ modes, the initial state $\Lambda^+_c$ is an isospin singlet, the final states ($\Delta^{++}$, $\Delta^+$, $\Delta^0$, $\Delta^-$) form an isospin quartet, ($K^+$, $K^0$) and ($\overline K^0$, $K^-$) form two isospin doublets.
There are two topologies contributing to the $\Lambda_c\to \Delta K$ modes, the $W$-exchange diagram $E^\prime$ and the color-suppressed internal $W$-emission diagram $C^\prime$, which are displayed in Fig.~\ref{topo}.
Topological decompositions of the $\Lambda_c\to \Delta K$ modes are
\begin{align}\label{amp1}
\mathcal{A}(\Lambda^+_c\to \Delta^{++}K^-)&=  -V_{cs}^{*}V_{ud}\,E^\prime_1,\\ \label{amp2} \mathcal{A}(\Lambda^+_c\to \Delta^{+}\overline K^0)&=\frac{1}{\sqrt{3}}V_{cs}^{*}V_{ud}\,E^\prime_2, \\\label{amp3}
 \mathcal{A}(\Lambda^+_c\to \Delta^{+}K^0)&= \frac{1}{\sqrt{3}}V_{cd}^{*}V_{us}\,C^\prime_1,\\\label{amp4}
 \mathcal{A}(\Lambda^+_c\to \Delta^{0}K^+)&=-\frac{1}{\sqrt{3}}V_{cd}^{*}V_{us}\,C^\prime_2.
\end{align}
The opposite sign in Eq.~\eqref{amp1} and Eq.~\eqref{amp2} is arisen from the quark components of kaons, $|K^-\rangle = -|s\bar u\rangle$ and $|\overline K^0\rangle = |s\bar d\rangle$.
And the opposite sign in Eq.~\eqref{amp3} and Eq.~\eqref{amp4} is from the quark component of $\Lambda^+_c$ baryon, $|\Lambda^+_c\rangle = |(ud-du)c\rangle$.
Notice the difference between $E^\prime_1$ and $E^\prime_2$ is $u\bar u$ or $d\bar d$ generated from vacuum.
And the difference between $C^\prime_1$ and $C^\prime_2$ is an exchange of $u\leftrightarrow d$ as spectator quarks.
If the isospin breaking is neglected, i.e., $u(\bar u)=d(\bar d)$, we get $E^\prime_1=E^\prime_2=E^\prime$ and $C^\prime_1=C^\prime_2=C^\prime$.
As a further illustration, we perform an isospin analysis for the $\Lambda_c\to \Delta K$ modes in Appendix~\ref{iso}.
The isospin relations between several $\Lambda_c\to \Delta K$ modes are supported by literatures such as Refs.~\cite{Savage:1989qr,Jia:2019zxi,Hsiao:2020iwc,Geng:2019awr}.

The magnitudes of decay amplitudes $\large|\mathcal{A}(\Lambda^+_c\to \Delta^{+}\overline K^0)\large|$ and $\large|\mathcal{A}(\Lambda^+_c\to \Delta^{+}K^0)\large|$ can be extracted from the branching fractions of $\Lambda^+_c\to \Delta^{++}K^-$ and $\Lambda^+_c\to \Delta^{0}K^+$ modes in the isospin symmetry,
\begin{align}\label{ext1}
&\left|\mathcal{A}(\Lambda^+_c\to \Delta^{+}\overline K^0)\right|^2=\frac{1}{3}\left|\mathcal{A}(\Lambda^+_c\to \Delta^{++}K^-)\right|^2, \\\label{ext2}
 &\left|\mathcal{A}(\Lambda^+_c\to \Delta^{+}K^0)\right|^2=\left|\mathcal{A}(\Lambda^+_c\to \Delta^{0}K^+)\right|^2.
\end{align}
Then the ratio $r\equiv\large|\mathcal{A}(\Lambda^+_c\to \Delta^{+} K^0)/\mathcal{A}(\Lambda^+_c\to \Delta^{+}\overline K^0)\large|$ is determined according to Eq.~\eqref{r}.
The magnitude of $\large|\mathcal{A}(\Lambda^+_c\to \Delta^{+}K^0)\large|$ can also be extracted from the branching fractions of $\Lambda^+_c\to \Delta^{++}K^-$ and $\Lambda^+_c\to \Delta^{+}K^0_{S,L}$ modes via following relation,
\begin{align}
  &\left|\mathcal{A}(\Lambda^+_c\to \Delta^{+}K^0)\right|^2=3\left|\mathcal{A}(\Lambda^+_c\to \Delta^{+}K^0_{S})\right|^2\nonumber\\&\quad\qquad+3\left|\mathcal{A}(\Lambda^+_c\to \Delta^{+}K^0_{L})\right|^2-\left|\mathcal{A}(\Lambda^+_c\to \Delta^{++}K^-)\right|^2.
\end{align}
In order to extract the strong phase $\delta$, we define the $K_{S}^{0}-K_{L}^{0}$ asymmetry in the $\Lambda^+_c\to \Delta^{+}K^0_{S,L}$ modes as
\begin{equation}\label{Rf}
   R(\Lambda^+_c,\Delta^+)\equiv\frac{\Gamma(\Lambda^+_c\to \Delta^{+}K^0_{S}) -\Gamma(\Lambda^+_c\to \Delta^{+}K^0_{L})}{\Gamma(\Lambda^+_c\to \Delta^{+}K^0_{S}) + \Gamma(\Lambda^+_c\to \Delta^{+}K^0_{L})},
 \end{equation}
which is derived to be \cite{Wang:2017ksn}
\begin{align}\label{Rfepsilon}
R(\Lambda^+_c,\Delta^+) \simeq -2\,r\cos\delta.
 \end{align}
The sub-leading terms are negligible since they are at order of $10^{-4}$, much smaller than the leading term which is $\mathcal{O}(10^{-2})$. If $R(\Lambda^+_c,\Delta^+)$ is measured, the strong phase $\delta$ is given by
\begin{align}
\delta=\pm arc\,cos\left(\left|R(\Lambda^+_c,\Delta^+)/2r\right|\right).
\end{align}
Since isospin symmetry is a very precise symmetry, the theoretical uncertainties of $r$ and $\delta$ can be well controlled.
The isospin symmetry in the charmed baryon decay can be examined by the $\Lambda^+_c \to \Sigma^+\pi^0$ and $\Lambda^+_c \to \Sigma^0\pi^+$ modes. The ratio of branching fractions of these two decay modes is \cite{PDG,BESIII:2015bjk}
\begin{align}
\left|\mathcal{A}(\Lambda^+_c \to \Sigma^+\pi^0)/\mathcal{A}(\Lambda^+_c \to \Sigma^0\pi^+)\right| = 0.97\pm 0.09,
\end{align}
which is in agreement with isospin prediction value. Isospin symmetry in the $\Lambda_c\to \Delta K$ modes is more reliable since it is not damaged by $\pi^0-\eta(\eta^\prime)$ mixing \cite{Gronau:2005pq,Feldmann:1998sh,Gardner:1998gz}.
Future measurements of branching fractions will be crucial to reduce uncertainties of the SM prediction of $CP$ asymmetry.

The branching fraction of $\Lambda^+_c\to \Delta^{++}K^-$ has be extracted by the partial wave analysis of $\Lambda^+_c\to pK^-\pi^+$ \cite{PDG,E791:1999ajq,Basile:1981hc,ACCMOR:1993gfl}. A more precise measurement is desirable.
According to Appendix~\ref{iso}, the ratio $\mathcal{B}r(\Delta^{0}\to p\pi^-)/\mathcal{B}r(\Delta^{0}\to n\pi^0)$ is $1/2$.
The branching fraction of $\Delta \to N\pi$ decay is $99.4\%$ \cite{PDG}.
Then the branching fraction of $\Lambda^+_c\to \Delta^{0}K^+$ can be extracted from the partial wave analysis of $\Lambda^+_c\to pK^+\pi^-$.
$\mathcal{B}r(\Lambda^+_c\to pK^+\pi^-)$ has been measured by Belle and LHCb experiments \cite{Belle:2015wxn,LHCb:2017xtf}.
Suffered from the large background, it is difficult to perform a partial wave analysis for $\Lambda^+_c\to pK^+\pi^-$ on Belle~(II).
The signal yield of $\Lambda^+_c\to pK^+\pi^-$ on LHCb at integrated luminosity of $1.0\,\rm {fb}^{-1}$ is $(392\pm 35)$ \cite{LHCb:2017xtf}.
LHCb will accumulate a data sample corresponding to a minimum of $300\,\rm {fb}^{-1}$ in the future \cite{LHCb:2018roe}.
The signal yield of $\Lambda^+_c\to pK^+\pi^-$ could reach to be $1.2\times 10^{5}$ and then the partial wave analysis is possible.
Besides, if the Super $\tau$-Charm facility (STCF) is constructed and operated, $\Lambda^+_c\to pK^+\pi^-$ can be analyzed on STCF. The expected number of $\Lambda_c\overline {\Lambda}_c$ on STCF is $5.6\times 10^{8}$ per year, which is $3$-orders higher than the total number produced on BEPCII \cite{BESIII:2015bjk,Lyu:2021tlb}.
The $\Lambda^+_c\to \Delta^{+}K^0_S$ mode can be constructed from the Dalitz plot of $\Lambda^+_c\to pK^0_S\pi^0$. $\mathcal{B}r(\Lambda^+_c\to pK^0_S\pi^0)$ was measured on BESIII with around $600$ events \cite{BESIII:2015bjk}.
According to the number of $\Lambda_c\overline {\Lambda}_c$, the signal yield of $\Lambda^+_c\to pK^0_S\pi^0$ could reach to be around $3\times 10^{6}$ per year on STCF.
The branching fraction of $\Lambda^+_c\to \Delta^{+}K^0_L$ is closed to $\Lambda^+_c\to \Delta^{+}K^0_S$. It could be extracted from $\Lambda^+_c\to pK^0_L\pi^0$ on STCF in the future since the detection efficiency of $K^0_L$  will be markedly improved \cite{Lyu}.

The tree interference induced $CP$ asymmetry in charm decay is highly sensitive to new physics due to the tiny weak phase $\phi$ in the SM.
Once the $CP$ asymmetry in the $\Lambda^+_c\to \Delta^{+}K(t)(\to \pi^+\pi^-)$ mode is confirmed by experiments, we can check if it is beyond the SM or not.
Compared to other charmed hadron decay or neutral meson mixing systems, the ambiguities from loop-induced quantities and $SU(3)_F$ breaking effects are avoided in the SM prediction.
Physics beyond the Standard Model would be confirmed indirectly by high-precision measurements without the possible mockery from low-energy QCD.
Even if no signal of new physics is observed in the $\Lambda^+_c\to \Delta^{+}K(t)(\to \pi^+\pi^-)$ mode, it could verify the $CP$-violating effect resulted from the interference between the CF and DCS amplitudes with the neutral kaon mixing predicted in the Standard Model.

\section{Summary}\label{sum}
The $CP$ asymmetry of $\Lambda_c^+$ decaying into $\Delta^+$ and neutral kaons can be extracted from the branching fractions of several $\Lambda_c\to \Delta K$ decays under the isospin symmetry.
It might be served as a clear hint of new physics in charm sector.

\section{Acknowledgement}
We are very grateful to Hsiang-Nan Li, Fu-Sheng Yu, Cheng-Ping Shen, Wei Shan and Shu-Lei Zhang for useful discussions.
This work was supported in part by the National Natural Science Foundation of China under Grants No. 12105099.

\appendix

\section{Isospin analysis}\label{iso}

In this appendix, we perform an isospin analysis of the $\Lambda_c\to \Delta K$ decays.
For the CF modes $\Lambda^+_c\to \Delta^{++}K^-$ and $\Lambda^+_c\to \Delta^{+}\overline K^0$, the weak Hamiltonian changes isospin as $\Delta I = 1$ and $\Delta I_3 = 1$.
Then we have an isospin amplitude $\mathcal{A}_{1,1}$ since
\begin{align}
\langle \Lambda^+_c|H_{\rm CF}=\langle0,0;1,1|=\langle1,1|.
\end{align}
For the $\Delta K$ system in the isospin limit, we have
\begin{align}
   |\Delta^{++}K^-\rangle &= \left|\frac{3}{2},\frac{3}{2};\frac{1}{2},-\frac{1}{2} \right\rangle = \frac{1}{2}|2,1\rangle+\frac{\sqrt{3}}{2}|1,1\rangle, \\
  |\Delta^{+}\overline K^0\rangle &= \left|\frac{3}{2},\frac{1}{2};\frac{1}{2},\frac{1}{2}\right\rangle = \frac{\sqrt{3}}{2}|2,1\rangle-\frac{1}{2}|1,1\rangle.
\end{align}
The isospin conservation requires the initial and final states have the same isospin.
The isospin amplitudes of the $\Lambda^+_c\to \Delta^{++}K^-$ and $\Lambda^+_c\to \Delta^{+}\overline K^0$ modes are  derived as
\begin{align}
\mathcal{A}(\Lambda^+_c\to \Delta^{++}K^-) & = \frac{\sqrt{3}}{2} \mathcal{A}_{1,1},\qquad \mathcal{A}(\Lambda^+_c\to \Delta^{+}\overline K^0) = -\frac{1}{2} \mathcal{A}_{1,1}.
\end{align}
By matching the topological decomposition, we get $\mathcal{A}_{1,1} = -2\,V_{cs}^{*}V_{ud}\,E^\prime/\sqrt{3}$ and Eq.~\eqref{ext1} is satisfied in the isospin symmetry.
For the DCS modes $\Lambda^+_c\to \Delta^{+}K^0$ and $\Lambda^+_c\to \Delta^{0}K^+$, the weak Hamiltonian changes isospin as $\Delta I = 1$ or $\Delta I = 0$, and $\Delta I_3 = 0$.
There are two isospin amplitudes $\mathcal{A}_{1,0}$ and $\mathcal{A}_{0,0}$ since
\begin{align}
\langle \Lambda^+_c|H_{\rm DCS}&=\langle0,0;1,0|+\langle0,0;0,0|=\langle1,0|+\langle0,0|.
\end{align}
For the $\Delta K$ final states, we have
\begin{align}
  & |\Delta^{+}K^0\rangle = \left|\frac{3}{2},\frac{1}{2};\frac{1}{2},-\frac{1}{2}\right\rangle = \frac{1}{\sqrt{2}}|2,0\rangle+\frac{1}{\sqrt{2}}|1,0\rangle, \\
 & |\Delta^{0} K^+\rangle = \left|\frac{3}{2},-\frac{1}{2};\frac{1}{2},\frac{1}{2}\right\rangle = \frac{1}{\sqrt{2}}|2,0\rangle+\frac{1}{\sqrt{2}}|1,0\rangle. \label{IsoDCS}
\end{align}
The isospin amplitudes of $\Lambda^+_c\to \Delta^{+}K^0$ and $\Lambda^+_c\to \Delta^{0}K^+$ are derived to be
\begin{align}
\mathcal{A}(\Lambda^+_c\to \Delta^{+}K^0) &= \frac{1}{\sqrt{2}} \mathcal{A}_{1,0},\qquad \mathcal{A}(\Lambda^+_c\to \Delta^{0}K^+) = \frac{1}{\sqrt{2}} \mathcal{A}_{1,0}.
\end{align}
Then Eq.~\eqref{ext2} is satisfied in the isospin symmetry.

For the strong decay $\Delta \to N\pi$, for instance $\Delta^0 \to p\pi^-$ and $\Delta^0 \to n\pi^0$, we can get the ratio between branching fractions of two decay modes by isospin analysis.
The strong Hamiltonian is isospin conserved.
For the $N\pi$ system,
\begin{align}
  & |p\pi^-\rangle = \left|\frac{1}{2},\frac{1}{2};1,-1\right\rangle = \frac{1}{\sqrt{3}}\left|\frac{3}{2},-\frac{1}{2}\right\rangle-\sqrt{\frac{2}{3}}\left|\frac{1}{2},-\frac{1}{2}\right\rangle, \\
 &  |n\pi^0\rangle = \left|\frac{1}{2},-\frac{1}{2};1,0\right\rangle = \sqrt{\frac{2}{3}}\left|\frac{3}{2},-\frac{1}{2}\right\rangle+ \frac{1}{\sqrt{3}}\left|\frac{1}{2},-\frac{1}{2}\right\rangle.
\end{align}
The isospin $(I,I_3)$  of $\Delta^0$ baryon is $(3/2,-1/2)$. Then the isospin amplitudes of $\Delta^0 \to p\pi^-$ and $\Delta^0 \to n\pi^0$ are
\begin{align}
\mathcal{A}(\Delta^0 \to p\pi^-) = \frac{1}{\sqrt{3}} \mathcal{A}_{\frac{3}{2},-\frac{1}{2}},\qquad \mathcal{A}(\Delta^0 \to n\pi^0) = \sqrt{\frac{2}{3}} \mathcal{A}_{\frac{3}{2},-\frac{1}{2}},
\end{align}
and $\mathcal{B}r(\Delta^{0}\to p\pi^-)/\mathcal{B}r(\Delta^{0}\to n\pi^0) = 1/2$. Similarly, the ratio $\mathcal{B}r(\Delta^+ \to n\pi^+)/\mathcal{B}r(\Delta^+ \to p\pi^0)$ is also $1/2$ based on the isospin analysis.



\end{document}